\documentclass[11pt]{iopart}
\usepackage{bm}
\usepackage{hyperref}
\usepackage{amssymb}
\usepackage{feynmp}
\usepackage{amsopn}
\usepackage{setstack}
\usepackage{slashed}
\usepackage{graphicx}
\RequirePackage{ifpdf}

\ifpdf
\else 
\fi

\newcommand{\Mp}{M_{\mathrm{P}}}
\newcommand{\Mew}{M_{\mathrm{EW}}}

\newcommand{\figlabel}[1]{\textsf{({#1})}}

\newcommand{\chargino}{\lambda}
\newcommand{\charginobar}{\bar{\lambda}}
\newcommand{\egauge}{e}
\newcommand{\coupling}{g}
\newcommand{\Leff}{L_{\mathrm{eff}}}
\newcommand{\sugroup}{\mathrm{SU}}
\newcommand{\ugroup}{\mathrm{U}}

\renewcommand{\d}{\mathrm{d}}

\newcommand{\im}{\mathrm{i}}

\makeatletter
\newcommand\numberwithin[2]{\@addtoreset{#1}{#2}}
\makeatother
\numberwithin{footnote}{section}

\begin{document}
\begin{fmffile}{diags}
	\begin{flushright}
		\textsf{DESY 09-163}
	\end{flushright}
	\title{Higgs production as a probe of Chameleon Dark Energy}
	
	\author{Philippe Brax$^{1}$, Clare Burrage$^2$, Anne-Christine Davis$^3$,
	\\ David Seery$^3$ and Amanda Weltman$^{3,4,5}$}
	\vspace{3mm}
	
	\address{$^1$ Institut de Physique Th\'{e}orique,
	  CEA, IPhT, CNRS, URA2306, F-91191 Gif-sur-Yvette c\'{e}dex,
	  France \\[3mm]
	$^2$ Theory Group,	 Deutsches Elektronen-Synchrotron DESY,	 D-22603,
	  Hamburg, Germany \\[3mm]
	$^3$ Department of Applied Mathematics and Theoretical Physics, \\
	  Centre for Mathematical Sciences, University of Cambridge,
	  Wilberforce Road, Cambridge, CB3 0WA, United Kingdom \\[3mm]
	$^4$ Department of Mathematics and Applied Mathematics,
	  University of Cape Town, Private Bag, Rondebosch, South Africa, 7700
	  \\[3mm]
	$^5$ Centre for Theoretical Cosmology Fellow}
	\vspace{3mm}

	\eads{\mailto{philippe.brax@cea.fr},
	\mailto{clare.burrage@desy.de},
	\mailto{A.C.Davis@damtp.cam.ac.uk},
	\mailto{djs61@cam.ac.uk},
	\mailto{amanda.weltman@uct.ac.za}}
	
	\begin{abstract}
		In this paper we study various particle physics effects of a light,
		scalar dark energy
		field with chameleon-like couplings to matter. We show that a
		chameleon model with only matter couplings will induce a
		coupling to photons. In doing so, we derive the first
		microphysical realization of a chameleonic dark energy model
		coupled to the electromagnetic field strength. This analysis
		provides additional motivation for current and near-future tests
		of axion-like and chameleon particles. We find a new bound on the
		coupling strength of chameleons in uniformly coupled models.
		We also study the effect of chameleon fields on Higgs production,
		which is
		relevant for hadron colliders.
		These are expected to manufacture Higgs
		particles through weak boson fusion, or associated production with
		a $Z$ or $W^\pm$.
		We show that, like the Tevatron, the LHC will not be able to rule
		out or observe chameleons through this mechanism,
		because gauge invariance of the low energy Lagrangian suppresses
		the corrections that may arise.
		 \vspace{3mm}
		\begin{flushleft}
			\textsf{%
			}
		\end{flushleft}
	\end{abstract}
	\maketitle

	\section{Introduction}
	\label{sec:introduction}

	We now have compelling evidence that the rate of 
	expansion of the universe is accelerating, requiring the universe to be
	dominated by an unknown form of matter, known as `dark energy,'
	characterized by the equation of state $p \approx - \rho$.
	This dark energy
	could be a cosmological constant
	unnaturally tuned at the
	level of 1 part in $10^{123}$ \cite{Weinberg:1988cp},
	or it may be associated with the
	potential energy of a scalar
	field
	\cite{Ratra:1987rm,Wetterich:1994bg,Zlatev:1998tr,Carroll:1998zi}.
	If a scalar field drives the accelerated expansion then it must be
	extremely light, with mass of order $H_0\sim 10^{-33}\mbox{ eV}$.
	The existence of light scalar fields results in new, long range
	``fifth forces,'' which are tightly constrained by experiment
	\cite{Will:2001mx}.	To avoid these constraints the energy scale
	controlling the coupling of such scalar fields to matter must
	be many orders of magnitude larger than the Planck scale
	\cite{Carroll:1998zi}.
	Explaining why this coupling is so weak is a major problem
	for dynamical dark energy models.

	These problems can be circumvented by allowing the scalar field mass
	mass to be determined by a \emph{chameleon} mechanism
	\cite{Khoury:2003aq,Khoury:2003rn,Mota:2003tm,Clifton:2004st,
	Mota:2006fz}. This
	makes the field heavy in a dense environment, but
	light in vacuum.  Such scalar fields hide from
	experimental searches for fifth forces
	\cite{Khoury:2003aq,Khoury:2003rn} and modifications of
	gravity \cite{Hu:2007nk,Brax:2008hh} in a novel way:
	in the interior of a massive object
	the chameleon is very heavy, and has a correspondingly short
	interaction length. An observer outside the body
	feels a scalar force sourced only by a thin shell of matter at
	the surface of the object. This suppresses unwanted fifth forces
	\cite{Khoury:2003aq,Khoury:2003rn}.
	Observations impose few
	constraints on the strength of the coupling of the scalar
	field to the Standard Model, allowing the
	energy scale of the coupling to be many orders of
	magnitude below the Planck scale \cite{Mota:2006fz}.

	Chameleonic scalar fields have been shown to have a
	successful classical phenomenology.	Their equation of state
	depends on the energy density of the universe.
	In the early universe
	the chameleon behaves as an additional matter component,
	but at late times it has an equation of state with roughly
	$w \approx -1$.
	This enables it to drive an era of accelerated expansion
	\cite{Brax:2004qh}.
	Weak constraints on the initial conditions ensure that the
	chameleon does not disrupt the dynamics of the early universe
	\cite{Brax:2004qh,Mota:2006fz}.
	In this paper we discuss these theories in
	the framework of quantum mechanics.
	This is of interest because the cosmological constant problem
	is essentially quantum mechanical.
	There is no profit in replacing the cosmological constant by some
	other theory which is equally unnatural once quantum corrections are
	taken into account.

 	The chameleon mechanism makes it mandatory for
	the dark energy scalar field to interact with conventional matter, and
	these interactions are potentially very strong.
	They may help us unravel the microphysics of
	dark energy if their indirect consequences could be measured
	\cite{Brax:2007tk,Weltman:2008fp,Weltman:2008ng,Burrage2009190,
	Brax:2009bk,Chou:2008gr,Upadhye:2009iv}.
	If the interactions are sufficiently strong, however, then
	an unambiguous \emph{direct} consequence may be observable:
	new, light scalar quanta must be present in the beam
	pipe of any particle accelerator, raising the
	prospect of observing dark energy in the laboratory.
	Fully consistent quantum mechanical realizations of the chameleon
	have not yet been constructed, and may suffer from the same naturalness
	and tuning difficulties as quintessence
	\cite{Kolda:1998wq,Peccei:2000rz,Doran:2002bc}.
	In this paper we do not attempt to address naturalness concerns,
	or the quantum-mechanical construction of the model.
	Instead, supposing such models to exist, we determine
	constraints which are imposed by experiment.

	Real or virtual chameleon-like particles will certainly be produced in
	particle collisions if dark energy couples to the electroweak gauge bosons.
 	In this paper, we will provide a microscopic derivation of
 	this coupling for any chameleon-like model.
	The models which furnish chameleon-like
	scalar particles at low energy are
	effective theories, valid below some energy scale $M_c$.
	In Ref.~\cite{Brax:2009aw},
	corrections to electroweak precision observables
	from particles with chameleonic or axionic couplings
	were analysed, without making a commitment to any specific choice of
	physics at energies above $M_c$. It was shown that very weak
	constraints on the energy scale of the chameleon coupling to
	matter ensure that the chameleon does not have an observable
	effect on measurements of the $Z$-width.
	For processes involving fermions and
	$\sugroup(2) \times \ugroup(1)$ gauge bosons,
	it was shown that
	large effects could always be absorbed into
	renormalizations of the Fermi constant, $G_F$, and the
	gauge boson masses.
	When Higgs processes are included
	it is no longer clear that
	large effects can be hidden in this way, potentially allowing
	Higgs production
	to function as
	a diagnostic of dark energy physics and its coupling to the
	Standard Model.

	In this paper, we focus on the consequences of such dark energy
	quanta for Higgs production. This is a key target for both the
	\emph{Large Hadron Collider} (LHC) and the
	\emph{Tevatron}.
	The details of Higgs production depend on which couplings occur in
	the theory.
	To achieve a successful chameleon
	phenomenology
	we assume conformal
	couplings to matter.%
		\footnote{All chameleon models developed to date have
				  conformal couplings to matter.  However there exist
				  related models of scalar fields with environment
				  dependent properties which do not require conformal
				  couplings \cite{Luty:2003vm,Nicolis:2008in}.}
	This has implications for
	interactions with gauge boson kinetic terms,
	which will be described in more detail in~\S\ref{sec:scalar-corrections}
	below.
	The Standard Model gauge bosons
	have conformally invariant kinetic terms
	and develop no couplings to a scalar field of this type.
	In this paper we show that
	violations of conformal symmetry, arising from couplings
	to matter species, can be communicated to the kinetic term
	by quantum corrections. Therefore,
	a coupling \emph{is} generated at energies below $M_c$
	via loops of charged heavy particles. This coupling has
	important phenomenological consequences which will be briefly
	recalled. In particular we will show that this coupling
	induces a coupling to the electromagnetic field and in so
	doing provides a motivation for the effects probed in quantum
	laser experiments
	\cite{Brax:2007hi,Brax:2007ak,Mirizzi:2005ng,Gies:2007su,
	Ahlers:2007st,Chou:2008gr,Upadhye:2009iv}.
	Our results apply for any light scalars
	with the requisite chameleon-like couplings, whatever their
	origin.

	What scale, $M_c$, should be associated with these new degrees of freedom?
	A conservative choice would be the GUT scale,
	$M_c \sim 10^{16}$ GeV, which may be connected
	with an early inflationary stage,
	or a seesaw explanation of neutrino mass
	\cite{Minkowski:1977sc,GellMann:1980vs,Yanagida:1979as}.
	If so, heavy neutrinos of mass $M_c$ might exist
	but would be sterile, having no
	$\sugroup(2) \times \ugroup(1)$ quantum numbers.
	This would not lead to the required coupling.
	Alternatively, if supersymmetry is realized in nature then
	$M_c$ could be associated with the scale at which it is
	spontaneously broken, perhaps of order $1$--$10$ TeV.
	In any supersymmetric completion of the Standard Model there exist
	fermionic partners of
	$\gamma$, $W^\pm$, $Z$ and the Higgs,
	known as gauginos and Higgsinos.
	(For a review see \cite{Binetruy:2006ad} and
	references therein.)
	The mass eigenstates of these particles (``charginos'') would naturally
	be of order $M_c$.
	In this article we remain agnostic about the nature	of whatever
	particles circulate within loops.
	We give our calculation in
	a form which can be specialized immediately to the minimal
	supersymmetric Standard Model (MSSM), but our
	conclusions are general and do not depend on the details of
	a specific implementation.
	In any case, the calculation we describe can be adapted easily
	to any heavy particle carrying the requisite quantum numbers.
	
	The outline of this paper is as follows.  In
	\S\ref{sec:axionic} we introduce our model and show that a
	coupling between gauge
	bosons and dark energy is generated at low energy by
	integrating out heavy particles. Some of the phenomenology associated
	with this coupling is presented. 	
	In \S\ref{sec:higgs} we briefly review Higgs production at a
	hadron collider, emphasizing the role of
	vertices between the electroweak gauge bosons and the Higgs.
	In \S\ref{sec:scalar-corrections} we compute corrections to the
	Higgs production rate arising from the low energy theory written down
	in \S\ref{sec:axionic}.
	We show that the effective Lagrangian
	includes new contact interactions between the gauge
	bosons and the Higgs. At scales smaller than $1/M_c$ these
	contact interactions
	resolve into heavy particle loops.
	We compute the correction to Higgs production
	from both effects.
	In \S\ref{sec:conclusions}
	we conclude by arguing that the effect of chameleon-like particles on
	Higgs production is expected to be rather small.
	
	Throughout this paper, we adopt units in which $c = \hbar = 1$.
	We set the reduced Planck mass, $\Mp \equiv (8 \pi G)^{-1/2}$,
	to unity. Our metric convention is $(-,+,+,+)$.

	\section{A low-energy theory of gauge bosons and scalars}
	\label{sec:axionic}

	Khoury \& Weltman
	\cite{Khoury:2003rn,Khoury:2003aq}
	suggested that a scalar field, $\chi$,
	might evade detection and yet remain
	light in vacuum if it
	coupled to the matter species, $\psi^i$,
	via a set of conformally rescaled metrics,%
		\footnote{If the different matter species do not interact,
		then each species can be chosen to couple to $\chi$
		through a single metric. In this case, the indices $i$ and $\alpha$
		are the same and the species $\psi^i$ couples to the metric
		$g_{ab}^i$ only. Where interactions among the matter species are
		present, each species may couple to more than one metric.
		
		Alternatively, if $\chi$ couples to matter through
		a species-independent
		conformal rescaling of the metric, then
		the strength of the coupling to each matter species
		is the same. For phenomenological purposes,
		however, we wish to relax this restriction. In the remainder
		of this paper we allow the matter couplings to be distinct, but
		frequently return to the minimal scenario where all
		coupling strengths are the same.}
	\begin{equation}
		g_{ab} \rightarrow g^\alpha_{ab} = f_\alpha(\beta_\alpha \chi) g_{ab} .
		\label{eq:metric}
	\end{equation}
	In this formula, $g_{ab}$ is the spacetime metric,
	and $\beta_\alpha \equiv 1/M_\alpha$ is a coupling scale that
	is not necessarily related to the cutoff controlling the
	validity of the theory, $M_c$.
	The Einstein frame action is
	\begin{equation}
		S = \int \d^4 x \; \sqrt{-g} \left(
			\frac{M_P^2}{2}R-\frac{1}{2}(\partial \chi)^2-V(\chi)
		\right)
  		-
  		\int \d^4 x \; \mathcal{L}_m(\psi^{i}, g^\alpha_{ab}),
	\end{equation}
	where $\mathcal{L}_m$ is the Lagrangian describing the dynamics of the
	matter fields, $\psi^i$.
	The dynamics of $\chi$ are controlled by an effective
	potential $V_{\mathrm{eff}}$, which depends explicitly on the
	environment through $\mathcal{L}_m$ and satisfies
	\begin{equation}
		V_{\mathrm{eff}} =
		V(\phi) + \frac{1}{2} \sum_\alpha
		f_\alpha^2(\beta_\alpha \chi)
		\frac{g_{ab}^\alpha}{\sqrt{-g^\alpha}}
		\frac{\delta \mathcal{L}_m (\psi^i, g_{ab}^\beta)}
		{\delta g_{ab}^\alpha} ,
	\end{equation}
	where the sum is over all metrics
	and $g^\alpha = \det g_{ab}^\alpha$.
	The conformal functions $f_\alpha$ and the potential $V$ must be chosen to
	realise a successful
	chameleon mechanism. This requires that the effective
	potential has a minimum.
	Another requirement is the existence
	of a \emph{thin shell} effect \cite{Khoury:2003rn}, which is realized
	if the scalar field potential and 
	coupling functions are chosen in such a way that large variations
	in the mass of the field can occur as the matter content of
	the local environment changes.
	Classical realizations of this idea have been constructed, but
	it is not known
	whether quantum realizations exist, or
	what condition the $f_\alpha$ must satisfy once quantum
	corrections are taken into account.
	In what follows
	we will neglect the indices $i$ and
	$\alpha$ for simplicity, but it is not
	necessary to assume that the dark energy field couples with
	the same strength to all matter fields.
	
	We work in the
	Einstein frame, in which the gravitational part of the action
	is that of general relativity.
	If the scalar field couples conformally with universal
	strength, then
	a classically equivalent
	description can be obtained by transforming to the Jordan frame
	in which all effects of the scalar field reappear in the
	gravitational sector.
	In the classical theory, physical observables are independent of
	this choice.
	In the quantum theory, however,
	a non-trivial Jacobian may be necessary to connect the Einstein-
	and Jordan-frame measures in the path integral
	\cite{Doran:2002bc}.
	In this paper we work in the Einstein frame
	from the outset, and neglect terms arising from the Jacobian.
	These make additional contributions to the coupling between
	gauge-boson kinetic terms and the dark energy scalar.
	We hope to return to this question in a future publication.

	Despite appearances, chameleonic couplings of the form
	\eref{eq:metric} do not necessarily
	give rise to large variations in fundamental constants or particle
	masses \cite{Brax:2004qh,Mota:2006fz}. This follows
	because the minimum of the effective potential
	is a cosmological attractor.
	The location of the minimum drifts only slowly, owing
	to time evolution of the matter density.
	This implies that variations in
	the scalar-dependent particle masses and fundamental constants are
	correspondingly small.

	The kinetic terms of spin-1 particles are conformally
	invariant, and are left inert under the substitution
	$g_{ab} \rightarrow g_{ab}^\alpha$ in Eq.~\eref{eq:metric}.
	However,
	particle physics is not conformal;
	any coupling to other matter species will generically break this
	invariance. If so, quantum corrections will cause the kinetic terms
	to depend on $g_{ab}^\alpha$.
	In this section we compute these threshold corrections, leading to an
	effective theory which describes the interaction of gauge bosons
	and the dark energy field at low energy.

	\subsection[Axion-like couplings from heavy particle loops]
		{Axion-like couplings from heavy particle loops%
		\footnote{We would like to thank D. Shaw for very helpful
		discussions while preparing the text of this section.}}
	\label{sec:charginos}

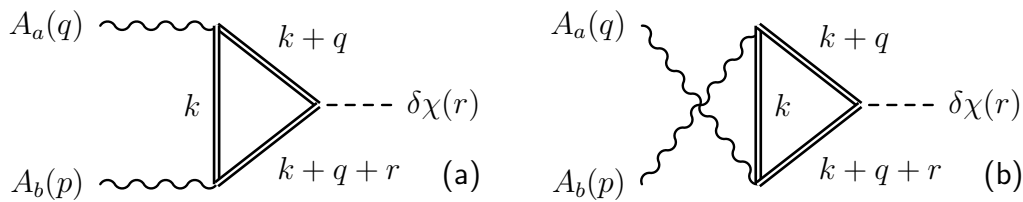
\begin{figure}
		\vspace{5mm}
		\hspace{15mm}
		\begin{fmfgraph*}(110,60)
			\fmfleft{l2,l1}
			\fmfright{r}
			\fmf{boson}{l1,v1}
			\fmf{boson}{l2,v2}
			\fmf{dbl_plain,label=$k$,label.side=left}{v2,v1}
			\fmf{dbl_plain,label=$k+q$,label.side=left}{v1,v3}
			\fmf{dbl_plain,label=$k+q+r$,label.side=left}{v3,v2}
			\fmf{dashes}{r,v3}
			\fmfv{label=$A_a(q)$,label.angle=180}{l1}
			\fmfv{label=$A_b(p)$,label.angle=180}{l2}
			\fmfv{label=$\delta\chi(r)$}{r}
			\fmffreeze
			\fmfforce{(0,0)}{l2}
			\fmfforce{(0,h)}{l1}
			\fmfforce{(w,0.5h)}{r}
			\fmfforce{(0.4w,h)}{v1}
			\fmfforce{(0.4w,0)}{v2}
			\fmfforce{(0.75w,0.5h)}{v3}
		\end{fmfgraph*}
		\hspace{4mm}
		\figlabel{a}
		\hfill
		\begin{fmfgraph*}(110,60)
			\fmfleft{l2,l1}
			\fmfright{r}
			\fmf{boson}{l1,v2}
			\fmf{boson}{l2,v1}
			\fmf{dbl_plain,label=$k$,label.side=right}{v2,v1}
			\fmf{dbl_plain,label=$k+q$,label.side=left}{v1,v3}
			\fmf{dbl_plain,label=$k+q+r$,label.side=left}{v3,v2}
			\fmf{dashes}{r,v3}
			\fmfv{label=$A_a(q)$,label.angle=180}{l1}
			\fmfv{label=$A_b(p)$,label.angle=180}{l2}
			\fmfv{label=$\delta\chi(r)$}{r}
			\fmffreeze
			\fmfforce{(0,0)}{l2}
			\fmfforce{(0,h)}{l1}
			\fmfforce{(w,0.5h)}{r}
			\fmfforce{(0.4w,h)}{v1}
			\fmfforce{(0.4w,0)}{v2}
			\fmfforce{(0.75w,0.5h)}{v3}
		\end{fmfgraph*}
		\hspace{4mm}
		\figlabel{b}
		\hspace{15mm}
		\mbox{}
		\caption{Diagrams contributing to the leading interaction
		between dark energy and the electroweak gauge bosons,
		which determine an effective operator acting on
		$A_a(q) A_b(p) \chi(r)$. Note that the momentum carried
		by $\chi$ is taken to flow into the diagram.
		Double lines represent a species of heavy fermion
		charged under $\sugroup(2) \times \ugroup(1)$.
		\label{fig:chi-diagrams}}
	\end{figure}

	The necessary corrections are depicted in Fig.~\ref{fig:chi-diagrams},
	in which a heavy fermionic particle circulates in the loop.
	Fermions couple to the metric~\eref{eq:metric} via a
	vielbein, $e^\mu_a$, which satisfies
	\begin{equation}
		g_{ab} = \eta_{\mu \nu} e^\mu_{(a} e^\nu_{b)} .
	\end{equation}
	There is an inverse vielbein, $e^a_\mu$, satisfying
	$e^\mu_a e^a_\nu = \delta^\mu_\nu$ and
	$e^\mu_a e^b_\mu = \delta^a_b$.
	The indices $\mu$,~$\nu$,~\ldots transform under a rigid
	Lorentz symmetry and can be coupled to Dirac $\gamma$-matrices.
	The action for a Dirac fermion, $\chargino$,
	with large mass $M$,
	can be written
	\begin{equation}
		L = \sqrt{-g}f^2( \beta \chi )
		\left\{
			- \charginobar ( \gamma^\mu e^a_\mu D_a + M ) \chargino
		\right\} ,
	\end{equation}
	where $D_a$ is a gauge-covariant derivative
	and $\charginobar \equiv \chargino^\dag \gamma^0$ is the spinor
	conjugate to $\chargino$.
	We suppose that $\chargino$ transforms under
	an Abelian symmetry
	with gauge coupling constant
	$\egauge$, so that
	\begin{equation}
		D_a \equiv
		\partial_a
		+
		\frac{1}{8}
		\gamma_{\mu\nu} \omega^{\mu\nu}_a
		-
		\im \egauge A_a ,
	\end{equation}
	where $\gamma_{\mu\nu} = [ \gamma_\mu, \gamma_\nu ]$.
	Note that the spin connexion, $\omega^{\mu\nu}_a$, transforms
	non-trivially under conformal rescalings.
	The calculation will be generalized to non-Abelian symmetries
	below Eq.~\eref{eq:chi-gauge-effective-operator}.
	Conformal invariance is broken by the fermion mass, $M$.
	Like the mass of any canonical field,
	$M$ varies under conformal transformations
	like $M \rightarrow f^{1/2}(\beta\chi)M$. If all energy
	scales ran in the same way with the vacuum expectation value
	of the scalar field then the effects of this coupling would
	never be observable, but this is not the case. Both
	the Planck scale, $M_P$, and the scale controlling the
	strength of the
	scalar coupling to matter, $\beta$, are unchanged by
	variations in the scalar field.

	We take
	the dark energy scalar to have
	a spatially-independent vacuum expectation value $\langle
	\chi \rangle = \bar{\chi}$, around which we quantize small fluctuations
	$\delta \chi$.
	(For this reason our calculation cannot be applied to very large
	spacetime volumes in which $\chi$ may develop appreciable gradients.)
	After rescaling the fermion fields to have
	canonical kinetic terms at leading order,
	the interaction between $\chargino$
	and these fluctuations can be written
	\begin{equation}
		\Leff \supseteq
		- \frac{1}{2}\sqrt{-g} \frac{\bar{f}'}{\bar{f}}
		\beta{\cal M}
		(\charginobar \chargino) \delta \chi ,
	\end{equation}
	where $\bar{f} \equiv f(\beta \bar{\chi})$
	and we have defined a conformally transformed
	mass ${\cal M} \equiv \bar{f}^{1/2}M$.
	Nevertheless, we emphasize that the physical mass of the fermion is $M$.
	A prime $'$ denotes the
	derivative of a function with respect to its argument.

	In each diagram of Fig.~\ref{fig:chi-diagrams},
	operators $A_a(q)$, $A_b(q)$
	and $\delta\chi(r)$ are inserted on the external legs, with all momenta
	flowing inwards.
	Fig.~\ref{fig:chi-diagrams}\figlabel{b}
	depicts the ``crossed'' diagram,
	which corresponds to reversing the sense of momentum flow in
	the fermion loop. It can be obtained from
	Fig.~\ref{fig:chi-diagrams}\figlabel{a}
	by the simultaneous replacements
	$p \leftrightarrow q$ and $a \leftrightarrow b$.
	Accounting for both diagrams
	we find that
	the correlation function $\langle A_a(q) A_b(p) \delta \chi(r) \rangle$
	can be written
	\begin{eqnarray}
		\fl\nonumber
		\langle A_a(q) A_b(p) \delta \chi(r) \rangle
		=
		- (2\pi)^4 \delta(p+q+r) \frac{ \bar{f}'}{\bar{f}}
		\frac{\egauge^2 \beta {\cal M}}{2} \cdot \delta\chi(r) \cdot A_a(q)
		A_b(p)\times \\
		\nonumber
		\hspace{-20mm} \mbox{}
		\int \frac{\d^4 k}{(2\pi)^4}
				\tr \left\{
			\gamma^a
			\frac{-\im (\slashed{k} + \slashed{q}) + {\cal M}}
				{(k+q)^2 + {\cal M}^2 - \im \epsilon}
			\frac{-\im (\slashed{k} - \slashed{p}) + {\cal M}}
				{(k-p)^2 + {\cal M}^2 - \im \epsilon}
			\gamma^b
			\frac{-\im \slashed{k} + {\cal M}}{k^2 +{\cal M}^2 - \im \epsilon}
		\right\} \\
		\hspace{-10mm} \mbox{} + \Big(
			\begin{array}{c}
				p \leftrightarrow q \\ a \leftrightarrow b
			\end{array}
		\Big) ,
		\label{eq:chi-gauge-correlation}
	\end{eqnarray}
	where `$\tr$' denotes a trace over Dirac indices.

	The $k$ integral is divergent, and
	such integrals are not guaranteed to be
	invariant under the rigid shift $k^a \rightarrow k^a + a^a$.
	In the absence of a shift symmetry, the integral can
	depend on the labeling of momenta within the loop.
	We apply dimensional regularization to maintain gauge invariance,
	after which
	it can be checked that the result is insensitive to
	the routing of momentum through the diagram.
	Eq.~\eref{eq:chi-gauge-correlation} also contains a potentially
	gauge-violating zero derivative term, which would be proportional
	to
	$\delta \chi A^a A_a$ in real space. This term vanishes when
	the $k$ integral is evaluated using any
	gauge-invariant regulator.
	Discarding terms which are proportional to the equations of
	motion,
	we find that Eq.~\eref{eq:chi-gauge-correlation} can be reproduced
	from an effective Lagrangian of the form
	\begin{equation}
		\Leff \supseteq
		\frac{\egauge^2 \beta}{96 \pi^2}
		\frac{\bar{f}'}{\bar{f}} \delta \chi
		F^{ab} F_{ab} + \Or (\partial^4) ,
		\label{eq:chi-gauge-effective-operator}
	\end{equation}
	where $\Or(\partial^4)$ denotes terms containing
	four or more derivatives.
	Eq.~\eref{eq:chi-gauge-effective-operator} applies for each species
	of heavy fermion in the theory.
	If there are many such fermions, each contributes
	with its own $\beta$ and $f$.
	If $\egauge$, $\bar{f}$ and $\bar{f}'$ are of order unity
	for all species,
	the coupling will be dominated by the fermion with largest $\beta$.
	Similar couplings would be induced by scalar particles, such as
	heavy sleptons. 

	The physical effect of the coupling in
	Eq.~\eref{eq:chi-gauge-effective-operator} could
	also be understood in the Jordan frame.
	There, one would typically neglect the effect of dark energy
	because of its tiny coupling, of order $\Mp^{-1}$.
	In the model we are considering, this neglect would be unjustified.
	The scalar field is more strongly coupled, and its effects
	would be
	manifest in the curvature of spacetime
	caused by particles participating in an interaction.
	The curvature scale would be associated with energies $M \ll \Mp$,
	which would be larger than that typically associated with
	gravitational phenomena.

	In this calculation, the gauge field $A_a$ was taken to be Abelian.
	However, it is clear that the same calculation generalizes
	immediately
	to non-Abelian fields
	for which the heavy fermion, $\lambda$,
	transforms in the fundamental representation.
	Consider any gauge group with generators
	$t_\alpha$, so that $A_a = A_a^\alpha t_\alpha$.
	The $t_\alpha$ may be normalized to satisfy
	\begin{equation}
		\Tr (t_\alpha t_\beta) = c_1 \delta_{\alpha \beta} ,
	\end{equation}
	for an arbitrary constant $c_1$,
	given that `$\Tr$' denotes a trace over indices in the gauge group.
	Eq.~\eref{eq:chi-gauge-effective-operator}
	therefore applies equally in a non-Abelian theory
	after the substitution $F^{ab} F_{ab} \rightarrow \Tr F^{ab} F_{ab}$.
	
	Eq.~\eref{eq:chi-gauge-effective-operator} will be accompanied
	by more complicated corrections which couple $\Tr F^{ab} F_{ab}$
	to all powers of $\delta \chi$.
	Resumming this expansion, we generate a coupling of the form
	$B(\beta_\gamma \chi) \Tr F^{ab} F_{ab}$
	for some function $B$
	and coupling scale $\beta_\gamma$.
	Eq.~\eref{eq:chi-gauge-effective-operator} allows us to estimate the
	scale $\beta_\gamma$,
	\begin{equation}
		\beta_\gamma \equiv \frac{\egauge^2 \beta}{96 \pi^2}
		.
		\label{bet}
	\end{equation}
	In the following we will assume, for simplicity,
	that no coupling to gluons is generated.
	The coupling to electromagnetism will give a small scalar
	field dependent contribution to the mass of atoms. If the theory were not
	of the chameleonic type, this would potentially lead to a strong violation
	of the weak equivalence principle. For chameleons, this will be heavily
	suppressed by the thin shell mechanism.
	
	\subsection{Constraints on Chameleon couplings}	

	Eq.~\eref{eq:chi-gauge-effective-operator}
	shows that, in a conformally-coupled theory,
	even if contact interactions involving $\Tr F^{ab} F_{ab}$
	are not present at high energies, they will inevitably be generated
	after passing the mass threshold of any matter species which
	couples both to $\chi$ and the gauge field.
	Therefore, laboratory and astrophysical bounds
	cannot be evaded merely by taking
	the $\delta \chi \cdot \Tr F^{ab} F_{ab}$ interaction to be absent,
	although their interpretation becomes model dependent.
	We believe this to be the first microphysical derivation
	of the coupling in Eq.~\eref{eq:chi-gauge-effective-operator}.

	This coupling means that chameleons are also axion-like particles:
	they couple to photons in an analogous way to the
	Peccei--Quinn axion, and therefore have a similar phenomenology. They
	are only `axion-like' because the mass and couplings of the field are not
	related as they are for	a standard axion, and because we are
	considering a scalar (rather than pseudo-scalar) field.  Observational
	constraints on the couplings of axion-like particles are very
	tight. However, applying these constraints to
	chameleonic fields is not straightforward because the mass of
	the chameleon field depends on its environment.	 The
	strongest constraints on axion-like particles come from their
	production in the cores of stars, but a chameleon field becomes very
	massive in the interior of the sun and therefore its production
	through scattering processes is suppressed. The constraints on
	chameleons from such observations are discussed in more detail in
	Ref.~\cite{Burrage:2008ii}.  For similar reasons
	chameleonic fields are not constrained by so-called
	``light shining through walls'' searches for axion-like
	particles \cite{PhysRevLett.59.759}.
	This is because the chameleon becomes heavy in the
	wall and is reflected by it rather than passing though.
	On the other hand,
	constraints that come from observing the behaviour of
	axion-like particles purely in diffuse environments can be applied to
	chameleonic fields.
	Strong constraints come from
	laboratory
	\cite{Gies:2007su,Ahlers:2007st,Chou:2008gr,Afanasev:2008jt}
	and astrophysical searches
	\cite{Brax:2007ak,Brax:2007hi,Burrage:2007ew,Burrage:2008ii,Davis:2009vk,
	Schelpe:2010he}.
	Indeed, it is possible that certain astronomical
	observations may be explained most simply by including
	light scalars which couple as chameleons
	\cite{Burrage:2009mj}.

	Under certain circumstances
	it is possible to translate the stringent bounds obtained from
	electromagnetic probes, discussed above,
	into bounds on the matter coupling.
	We assume a minimal model in which the dark energy couples to matter
	with a uniform strength $\beta$, irrespective of species.
	This coupling is subject to only
	mild restrictions, depending on the precise self-interaction potential
	which is chosen for the chameleon field. Even where such restrictions
	exist, they typically require $\beta$ to be no smaller than the
	ordinary scale of nuclear physics, $\beta \lesssim (100\;\mbox{GeV})^{-1}$
	\cite{Mota:2006fz}.
	For $\beta \bar{\chi} \lesssim 1$ and gauge coupling
	$\egauge \approx 0.5$, which is roughly the scale of the
	$\sugroup(2)$ and $\ugroup(1)$ couplings of the Standard Model,
	we find
	\begin{equation}
		\beta_\gamma \approx 10^{-4} \beta .
		\label{eq:new-bound}
	\end{equation}
	Therefore,
	constraints on $\beta_\gamma$ can be translated to
	limits on the matter coupling, $\beta$.
	Unfortunately this constraint is highly model-dependent, and can be
	weakened arbitrarily by decreasing the gauge coupling
	$\egauge$. Strong constraints are obtained only when
	$\egauge$
	can be determined by other means.

	The strongest bound on $\beta_\gamma$ follows from
	observations of the polarization of starlight in the Milky Way,
	yielding $\beta_\gamma \lesssim (10^{9}\;\mbox{GeV})^{-1}$
	for models in which the mass of the field satisfies
	$m_{\chi}\lesssim 10^{-11}\mbox{ eV}$ in the interstellar medium
	\cite{Burrage:2008ii}.%
		\footnote{Constraints on the coupling of
		dark energy to photons,  $\beta_\gamma \lesssim
		(10^6\;\mbox{GeV})^{-1}$, also come from the PVLAS
		\cite{Brax:2007hi} and GammeV \cite{Chou:2008gr} experiments,
		and apply to models where the
		field is sufficiently light in the vacuum tube and
		sufficiently heavy in the walls of the tube to prevent it
		escaping from the experiment.}
	In models where Eq.~\eref{eq:new-bound} applies,
	it follows that there is a new, stronger
	bound on the matter coupling,%
		\footnote{Consistency with our assumptions requires
		$\beta \bar{\chi} \lesssim 1$, corresponding to
		$\bar{\chi} \lesssim \Mp/10^{10}$,
		which should be easily satisfied for the chameleon field in all
		relevant backgrounds.}
	\begin{equation}
		\beta \lesssim \frac{1}{10^5\;\mbox{GeV}}
		\simeq \frac{10^{14}}{\Mp} .
	\end{equation}

	If the strength of the coupling of the scalar field to the gauge
	bosons is strong, we might also expect to see the scalar field in
	particle colliders.
	Because the chameleonic
	field is light in the vacuum of a particle collider it cannot be
	integrated out of the theory.
	In Ref.~\cite{Brax:2009aw} the
	effects of dark energy on precision electroweak observables were
	discussed. Corrections from processes with
	scalar fields in the final state were shown to contribute only if the
	scale of the coupling is low. The best constraints come from
	observations of the width for $Z$-decay, and corrections to this from the
	scalar field are invisible if $\beta_{\gamma}\lesssim (10^{2}\mbox{
	GeV})^{-1}$.
	It was also shown that large corrections due to the scalar
	field are screened in all $2 \rightarrow 2'$
	fermion scattering interactions due
	to a combination of gauge invariance and the structure of boson/lepton
	couplings in the Standard Model. The only processes in the electroweak
	sector to which this screening theorem does not apply are those
	involving Higgs bosons.
	For this reason, we might wonder whether coupling a
	dark energy scalar to the Higgs boson would lead to a large
	enhancement of Higgs production in particle colliders.
	In the next section we will show that this loophole can be closed,
	and that one can expect all electroweak processes to be screened
	from chameleon corrections.

	\section{Higgs production}
	\label{sec:higgs}

	\subsection{Production at particle colliders}
	\label{sec:channels}
	
	Below the electroweak symmetry breaking scale
	of order 1 TeV,
	interactions of the lightest neutral
	Higgs, $h$, with the $Z$ boson are described by
	cubic and quartic couplings \cite{Weinberg:1967tq},
	\begin{equation}
		L_{ZZh} = \int \d^4 x \; Z_a Z^a \left\{
			\sqrt{ 2^{1/2} G_F} M_Z^2 h + \sqrt{2} G_F M_Z^2 h^2
		\right\} ,
		\label{eq:zzh}
	\end{equation}
	The
	$Z$ mass is	$M_Z \simeq 91.2$ GeV and the Fermi constant,
	$G_F$, is measured experimentally to be
	$G_F \simeq 1.17 \times 10^{-5}$ GeV$^{-2}$ \cite{Amsler:2008zzb}.
	Eq.~\eref{eq:zzh} is numerically correct for a minimal Standard Model
	Higgs. In a two Higgs doublet model these couplings are shared
	between the two neutral scalars,
	leading to suppression by a numerical factor.%
		\footnote{In a supersymmetric Standard Model
		this factor is $\sin(\beta - \alpha)$, where $\beta$
		parametrizes the ratio of Higgs vacuum expectation values,
		$\tan \beta \equiv v_2 / v_1$,
		and $\alpha$
		is an angle which occurs when diagonalizing the Higgs
		mass matrix.
		See, eg., Ref.~\cite{Binetruy:2006ad}.
		In a simple MSSM, $\sin (\beta - \alpha)$ may be near unity.}
	Analogous interactions for the $W^\pm$
	are obtained by the substitutions
	$Z_a Z^a \rightarrow 2 W^+_a W^{- a}$
	and $M_Z \rightarrow M_W$.

	A summary of the methods for Higgs production at a hadron collider
	can be found in the recent book by Kilian \cite{Kilian:2003pc}.
	(See also Ref.~\cite{Gunion:1989we}.)
	\begin{figure}
		\vspace{5mm}
		\mbox{}
		\hfill
		\begin{fmfgraph*}(80,60)
			\fmfleft{l1,l2}
			\fmfright{r1,r2,r}
			\fmf{fermion}{l1,vt,r1}
			\fmf{fermion}{r2,vb,l2}
			\fmf{boson,label=$Z,, W^\pm$,label.side=right}{vt,v}
			\fmf{boson,label=$Z,, W^\pm$,label.side=left}{vb,v}
			\fmf{dashes}{v,r}
			\fmffreeze
			\fmfforce{nw}{l1}
			\fmfforce{ne}{r1}
			\fmfforce{se}{r2}
			\fmfforce{sw}{l2}
			\fmfforce{(w,0.5h)}{r}
			\fmfforce{(0.5w,0.5h)}{v}
			\fmfforce{(0.5w,h)}{vt}
			\fmfforce{(0.5w,0)}{vb}
			\fmfv{label=$q$}{l1}
			\fmfv{label=$q$}{r1}
			\fmfv{label=$\bar{q}$}{l2}
			\fmfv{label=$\bar{q}$}{r2}
			\fmfv{label=$h$}{r}
		\end{fmfgraph*}
		\hspace{5mm}
		\figlabel{a}
		\hfill
		\begin{fmfgraph*}(80,60)
			\fmfleft{l1,l2}
			\fmfright{r1,r2}
			\fmf{fermion}{l1,v1}
			\fmf{fermion}{l2,v1}
			\fmf{boson,label=$W^\ast$}{v1,v2}
			\fmf{boson}{v2,r1}
			\fmf{dashes}{v2,r2}
			\fmfv{label=$q'$}{l1}
			\fmfv{label=$\bar{q}$}{l2}
			\fmfv{label=$W$}{r1}
			\fmfv{label=$h$}{r2}
		\end{fmfgraph*}
		\hspace{5mm}
		\figlabel{b}
		\hfill
		\mbox{}
		\vspace{5mm}
		\caption{\label{fig:higgs-production}Higgs production channels.
		In~\figlabel{a}, weak boson fusion is the next-to-leading
		process at a hadron collider, but benefits from
		accurate background subtraction.
		A $q \bar{q}$ pair undergoes the splitting
		$q \bar{q} \rightarrow q \bar{q} V \bar{V}$, where $q$ is
		a generic quark species and $V$ is a vector boson.
		The final state Higgs is radiated via fusion
		of the intermediate $V \bar{V}$ pair.
		In~\figlabel{b}, associated production occurs when
		two quarks, $q'$ and $\bar{q}$, fuse to form an off-shell $W$.
		The final state is achieved by Higgsstrahlung radiation.
		At a lepton collider the initial state $q'$, $\bar{q}$ can
		be replaced by an $e^+ e^-$ pair and an intermediate
		$Z^\ast$. The final state Higgs is produced in association
		with an on-shell $Z$.}
	\end{figure}
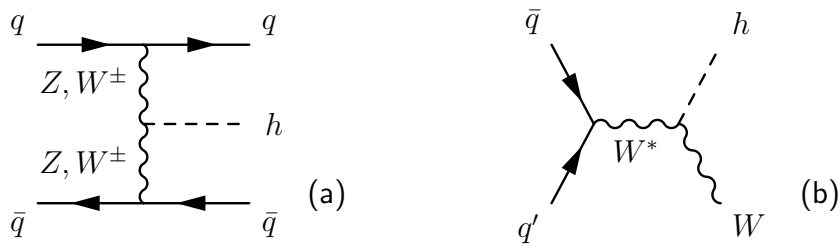
	One especially important production mechanism is
	`weak boson fusion,'
	$V \bar{V} \rightarrow h$,
	shown in
	Fig.~\ref{fig:higgs-production}\figlabel{a},
	where $V$ is any of the vector bosons $W^\pm, Z$.
	This process is generally sub-dominant
	to the rate of gluon--gluon fusion, but benefits from more
	accurate background subtraction.
	The precursor bosons originate within spectator quarks,
	as shown in Fig.~\ref{fig:higgs-production}\figlabel{a}
	and discussed in more detail in \S\ref{sec:effective-w}
	below.
	Following a boson fusion event, the spectator quarks are disrupted
	and initiate transverse hadronic jets.
	
	Another interesting mechanism is `associated production,'
	shown in Fig.~\ref{fig:higgs-production}\figlabel{b},
	in which two quarks fuse to form a 
	$W$ resonance that subsequently decays
	via ``Higgsstrahlung'' into an on-shell $W$ and a Higgs,
	$q' \bar{q} \rightarrow W^\ast \rightarrow W h$.
	Then its primary decays will be to $b \bar{b}$ pairs.
	Weak boson fusion and associated production rely on
	couplings of the Higgs to two vector bosons.
	These vertices are present in the Standard Model and in many
	theories of beyond-the-Standard-Model physics,
	although in certain cases they can be tuned to be absent.
	In this section, we study the possibility that corrections due to
	dark energy can change the relationship of these vertices to the
	other measurable parameters of the Lagrangian.
	Our analysis applies for a large class of models in which these
	two-boson couplings are represented by
	Eq.~\eref{eq:zzh} and its generalization to $W^+ W^-$
	interactions.
	
	\subsection{The effective $W$ approximation.}
	\label{sec:effective-w}
	
	The processes of weak boson fusion and
	associated production
	can be studied using conventional perturbation
	theory, but alternative methods exist. An especially useful
	tool is the
	\emph{effective $W$ approximation}
	(see, eg., Ref.~\cite{Kilian:2003pc}).
	For weak boson fusion, this means that the
	vector boson precursors
	are taken to be on-shell partons within colliding hadrons.
	In the parton picture, a hadron consists of three valence quarks
	which are surrounded by a sea of virtual particles. A probe which
	samples this sea at sufficiently high resolution has a chance
	to resolve the virtual quanta, rather than
	valence quarks. Since quarks participate in the electroweak interaction,
	$Z$ and $W^\pm$ bosons will be found within the virtual sea.
	Its precise composition can be determined by solving
	a system of DGLAP-like equations, which can be thought of as an
	approximate Boltzmann hierarchy \cite{Collins:1988wj}.
	In this picture, calculations involving hadron collisions with
	vector boson intermediate states simplify considerably.
	In the remainder of this paper
	we will work within the
	parton picture
	and the effective $W$ approximation.
	
	The utility of this description
	is that a complex cross-section can be factorized
	into a sequence of simpler subprocesses
	\cite{Lindfors:1985yp,Johnson:1987tj}.
	Dropping the contribution from $Z$ bosons,
	the effective cross-section for weak boson fusion can be
	written \cite{Gunion:1989we,Kilian:2003pc}
	\begin{equation}
		\sigma_{\mathrm{eff}} =
		\int \frac{\d x_1}{x_1} \; \frac{\d x_2}{x_2}
		\sum_{\lambda \in { \pm, L }}
		\Gamma( W^+_\lambda W^-_\lambda \rightarrow h )_{x_1 x_2 s}
		F_{W^+_\lambda}(x_1) F_{W^-_\lambda}(x_2) ,
		\label{eq:effective-w}
	\end{equation}
	where ${ \pm, L }$ are transverse and longitudinal polarization
	modes, and $s$ is the centre of momentum energy of the collision.
	The functions $F(x)$
	can be calculated or measured experimentally.
	Eq.~\eref{eq:effective-w} shows that the remaining chameleon
	corrections can be determined from the
	on-shell rate $\Gamma( W^+ W^- \rightarrow h )$,
	taken to occur at centre of momentum energy $x_1 x_2 s$.
	In what follows, we will concentrate on modifications to this
	quantity.
	
	\subsection{New physics and the $W^+ W^- h$, $ZZh$	 vertices}

	The rate $\Gamma( V \bar{V} \rightarrow h )$
	depends on the amplitude
	of a Green's function describing effective
	$ZZH$ and $WWH$
	vertices.
	This Green's function
	captures the appearance of new physics in Eq.~\eref{eq:effective-w}.
	In many theories of
	physics beyond the Standard Model,
	the form of the cubic and quartic vertices in
	Eq.~\eref{eq:zzh} are unmodified.
	In these theories,
	corrections to the self-energy diagrams of $Z$, $W^\pm$, $\gamma$
	and $H$ particles disrupt tree-level relationships between
	the particle masses $\{ M_Z, M_W, M_H \}$,
	the fine structure constant $\alpha$,
	and the Fermi constant $G_F$.
	This disruption can be summarized using
	\emph{oblique parameters}
	introduced by Peskin \& Takeuchi
	\cite{Peskin:1990zt,Peskin:1991sw}
	and refined by Maksymyk, London \& Burgess
	\cite{Maksymyk:1993zm,Burgess:1993mg}.
	In our model there are both oblique and non-oblique
	corrections.
	To summarize our results
	we use the oblique notation of Refs.~\cite{Maksymyk:1993zm,Burgess:1993mg}
	and note differences explicitly where they occur.

	Consider the Green's function
	describing an effective $ZZh$ vertex, from which
	the $WWh$ result can be derived after trivial modifications.
	The S-matrix element must depend on
	\vspace{5mm}
	\begin{equation}
		\left|
		\parbox{23mm}{\begin{fmfgraph*}(40,40)
			\fmftop{t}
			\fmfbottom{b}
			\fmfright{r}
			\fmf{boson}{t,v}
			\fmf{boson}{b,v}
			\fmf{dashes}{v,r}
			\fmfv{label=$Z$}{t}
			\fmfv{label=$Z$}{b}
			\fmfv{label=$h$}{r}
		\end{fmfgraph*}}
		\right|^2
		\propto
		M_Z^4 G_F \times \mbox{wavefunction renormalizations} ,
		\label{eq:fusion}
	\end{equation}
	\vspace{5mm}
	where the wavefunction renormalizations arise on transition to the
	$S$-matrix.
	We work in unitary gauge, where the 
	Goldstone modes associated
	with the $\sugroup(2)$ Higgs doublet are absorbed as longitudinal
	polarizations of $Z$ and $W^\pm$.
	Accounting for oblique corrections, the $Z$ propagator is
	\begin{equation}
		\langle Z^a(k_1) Z^b(k_2) \rangle
		= -\im (2\pi)^4 \delta(k_1 + k_2)
			\left( \eta^{ab} + \frac{k^a k^b}{M_Z^2} \right)
		\Delta'(k^2) ,
		\label{eq:Z-propagate}
	\end{equation}
	where $\Delta'(k^2)$ takes the form
	\begin{equation}
		\Delta'(k^2)^{-1} \equiv k^2 + M_Z^2 - \Pi_{ZZ}(k^2) .
		\label{eq:Z-delta}
	\end{equation}
	In Eqs.~\eref{eq:Z-propagate}, $k$ stands for either $k_1$ or $k_2$.
	The quantity $\Pi_{ZZ}$ is defined as follows.
	We choose
	$\im \Pi^{ab}_{ZZ}(k^2)$ to be
	the sum of all one-particle-irreducible graphs connecting an ingoing
	and an outgoing $Z$.
	In vacuum this has a unique tensorial decomposition,
	\begin{equation}
		\Pi^{ab}_{ZZ}(k^2) \equiv \eta^{ab} \Pi_{ZZ}^{(0)}(k^2)
		+ k^a k^b \Pi_{ZZ}^{(2)}(k^2) .
	\end{equation}
	We neglect the term involving $\Pi_{ZZ}^{(2)}$
	and can therefore drop superscript `$0$'s without
	ambiguity,
	so that $\Pi^{(0)}_{ZZ} \rightarrow \Pi_{ZZ}$.
	It is this quantity which appears in Eq.~\eref{eq:Z-delta}.
	If the mass of external fermions is at most $\sim M_f$,
	this neglect is equivalent to dropping powers of $M_f/M_Z$.
	It is likely to be a good approximation
	provided the Higgs is not too heavy:
	for a Higgs lighter than the top mass, $M_t \simeq 173$ GeV,
	decay into a top quark
	is kinematically forbidden.
	Therefore, $M_f/M$ is at most of order $10^{-1}$ to $10^{-2}$.
	
	Including oblique corrections, the $Z$ mass becomes
	\begin{equation}
		M_Z^2 = \tilde{M}_Z^2 \left( 1 - \frac{\Pi_{ZZ}(-M_Z^2)}{M_Z^2}
		\right) .
	\end{equation}
	A similar formula
	can be written for the $W$ propagator,
	making the replacements $M_Z \rightarrow M_W$ and
	$\Pi_{ZZ} \rightarrow \Pi_{WW}$.
	Quantities with a tilde, such as $\tilde{M}_Z$
	and $\tilde{G}_F$, refer to the value of these parameters in
	the absence of oblique corrections.
	With the same conventions,
	the Fermi constant
	satisfies
	\begin{equation}
		G_F = \tilde{G}_F \left( 1 + \frac{\Pi_{WW}(0)}{M_W^2} \right) .
	\end{equation}
	$G_F$ parametrizes the strength of the weak force near zero momentum
	transfer.
	Allowing for these shifts,
	the decay rate $\Gamma(ZZ \rightarrow h)$
	is related to the pure Standard Model rate by the rule
	\begin{eqnarray}
		\fl\nonumber
		\frac{\Gamma(ZZ \rightarrow h)}{\tilde{\Gamma}(ZZ \rightarrow h)}
		=
		1 + 2 \frac{\Pi_{ZZ}(-M_Z^2)}{M_Z^2} - \frac{\Pi_{WW}(0)}{M_W^2}
		+ 2 \Pi_{ZZ}'(-M_Z^2) + \Pi_{HH}'(-M_H^2) \\
		= 1 + \alpha ( 2 V + R ) .
		\label{eq:decay-rateZ}
	\end{eqnarray}
	For $W$ bosons the decay rate is
	\begin{eqnarray}
		\fl\nonumber
		\frac{\Gamma(WW \rightarrow h)}{\tilde{\Gamma}(WW \rightarrow h)}
		=
		1 + 2 \frac{\Pi_{WW}(-M_W^2)}{M_W^2} - \frac{\Pi_{WW}(0)}{M_W^2}
		+ 2 \Pi_{WW}'(-M_W^2) + \Pi_{HH}'(-M_H^2) \\
		= 1 + \alpha ( 2 W + R ) ,
		\label{eq:decay-rateW}
	\end{eqnarray}
	where, as above, $\alpha \approx 1/137$ is the fine structure constant.
	Eqs.~\eref{eq:decay-rateZ} and~\eref{eq:decay-rateW} have been written
	in terms of the conventional oblique quantities
	$V$ and $W$, which are defined to satisfy \cite{Maksymyk:1993zm}
	\begin{eqnarray}
		\alpha V & \equiv
			\frac{\d}{\d k^2} \left. \Pi_{ZZ}(k^2)
			\right|_{k^2 = - M_Z^2} - \frac{\Pi_{ZZ}(0) - \Pi_{ZZ}(-M_Z^2)}
			{M_Z^2} , \\
		\alpha W & \equiv
			\frac{\d}{\d k^2} \left. \Pi_{WW}(k^2)
			\right|_{k^2 = - M_W^2} - \frac{\Pi_{WW}(0) - \Pi_{WW}(-M_W^2)}
			{M_W^2} .
	\end{eqnarray}
	In addition, we have introduced a new quantity $R$
	which is a measure
	of the Higgs' wavefunction renormalization,
	\begin{equation}
		\alpha R \equiv
			\frac{\d}{\d k^2} \left. \Pi_{HH}(k^2) \right|_{k^2 = -M_H^2}
			+ \frac{\Pi_{ZZ}(0)}{M_Z^2} .
	\end{equation}
	If the dark energy coupling scale is greater than
	the typical scale of electroweak processes,
	$\Mew \sim 1 \, \mbox{TeV}$,
	we expect
	$V$ and $W$ to be negligible \cite{Brax:2009aw}.
	The impact of new physics is therefore contained entirely in $R$.
	
	\section{Corrections from a dark energy scalar}
	\label{sec:scalar-corrections}

	Eqs.~\eref{eq:decay-rateZ}--\eref{eq:decay-rateW}
	determine the sensitivity of weak boson fusion and
	Higgsstrahlung
	to new physics.
	This sensitivity is measured by
	the Higgs oblique parameter, $R$.
	In this section we make a quantitative estimate of its magnitude.
	To do so, we must be precise about the
	corrections $\Pi_{ZZ}$ and $\Pi_{WW}$ which
	modify the Standard Model prediction.
	In \S\ref{sec:low-energy-oblique} we
	determine these quantities in a low energy
	chameleon-type model coupled to the gauge bosons.
	We calculate oblique corrections to the
	production rate, and show that they are
	sensitive to the high energy completion
	of the theory.
	In \S\ref{sec:charginos-higgs} we compute
	non-oblique corrections generated by
	integrating out heavy fermions.
	These are
	described by a new quartic coupling between
	the Higgs field and the gauge bosons.

	\subsection{Oblique corrections in the low-energy theory}
	\label{sec:low-energy-oblique}

	A dark energy
	field induces both straight and oblique corrections to
	the vacuum polarizations of the Higgs and gauge bosons.
	In Ref.~\cite{Brax:2009aw} it was argued that
	the straight corrections effectively
	divide into processes involving ``chameleonstrahlung,'' where
	dark energy particles are produced but escape the detector,
	and a collection of ``bridges,'' ``daisies'' and ``rainbows'' which
	dress the bare processes of the Standard Model.
	At leading order, these dressings are momentum independent.
	Chameleonstrahlung was shown to
	give constraints roughly comparable to those arising from oblique
	corrections. In this paper we focus on oblique corrections only.
	
	After electroweak symmetry breaking, we can parameterize the
	interactions of Eq.~\eref{eq:chi-gauge-effective-operator}
	by adopting an effective $Z$ boson Lagrangian of the form
	employed in Ref.~\cite{Brax:2009aw},
	\begin{equation}
		\fl
		S = - \frac{1}{4} \int \d^4 x \;
			\left\{
				B(\beta \chi) (\partial^a Z^b - \partial^b Z^a)
				(\partial_a Z_b - \partial_b Z_a)
				+ 2 B_H(\beta_H \chi) M_Z^2 Z^a Z_a
			\right\} .
		\label{eq:Z-coupling}
	\end{equation}
	
	The functions
	$B$ and $B_H$ should satisfy $B(0) = B_H(0) = 1$, but
	depend on the details of ultra-violet physics.
	More precisely, they
	are derived from Eq.~\eref{eq:chi-gauge-effective-operator}
	and similar higher-order diagrams involving more powers of
	$\delta \chi$.
	Likewise,
	the couplings $\beta$ and $\beta_H$ are inherited from whatever
	heavy particles are integrated out to generate this interaction.
	Working with a sharp momentum cutoff,
	the $Z$ vacuum polarization was determined in
	Ref.~\cite{Brax:2009aw}
	and found to be
	\begin{eqnarray}
		\fl\nonumber
		\Pi_{ZZ}(k^2) = \frac{\beta^2}{8\pi^2}
			\frac{\bar{B}^{\prime 2}}{\bar{B}}
			\int_0^1 \d x \;\Bigg\{ \frac{2 k^2 + \gamma^2 M_Z^2}{4}
				\left[
					\Lambda^2 - \frac{\Lambda^2}{2}
						\frac{\Lambda^2}{\Lambda^2 + \Sigma_Z^2}
					- \Sigma_Z^2 \ln \left(
						1 + \frac{\Lambda^2}{\Sigma_Z^2}
					\right)
				\right]
				\\ \nonumber
				\mbox{} +
				(x k^2 + \gamma M_Z^2)^2
				\left[
					- \frac{1}{2} \frac{\Lambda^2}{\Lambda^2 + \Sigma_Z^2}
					+ \frac{1}{2} \ln \left(
						1 + \frac{\Lambda^2}{\Sigma_Z^2}
					\right)
				\right]
				\\
				\mbox{} -
				\frac{\Omega}{2} (k^2 + \epsilon M_Z^2)
				\left[
					\frac{\Lambda^2}{2} - \frac{M_\chi^2}{2}
					\ln \left(1 + \frac{\Lambda^2}{M_\chi^2} \right)
				\right]
			\Bigg\} ,
			\label{eq:self-energy}
	\end{eqnarray}
	where $\Omega$ satisfies
	\begin{equation}
		\Omega \equiv \frac{\bar{B}'' \bar{B}}{\bar{B}^{\prime 2}}
	\end{equation}
	and $\bar{B} \equiv B(\beta \bar{\chi})$.
	The parameters $\epsilon$ and $\gamma$ are defined by
	\begin{eqnarray}
		\epsilon & = \frac{B_H''}{B''} \frac{\beta_H^2}{\beta^2} \\
		\gamma & = \frac{B_H'}{B'} \frac{\beta_H}{\beta} .
	\end{eqnarray}
	Also, $\Sigma_Z^2$ represents
	\begin{equation}
		\Sigma_Z^2 \equiv x(1-x) k^2 + (1-x) M_Z^2 + x M_\chi^2 ,
		\label{eq:SigmaZ}
	\end{equation}
	where $M_\chi$ is the mass of the dark energy fluctuation
	$\delta \chi$.
	Near $k^2 \approx 0$, $\Pi_{ZZ}(k^2)$ has the approximate form
	\begin{equation}
		\Pi_{ZZ}(k^2) \approx
		\frac{\beta_H^2 \Lambda^2}{32\pi^2}
		\frac{\bar{B}_H^{\prime 2}}{\bar{B}}
		\left( \frac{1}{2} - \frac{\bar{B}_H'' \bar{B}}{\bar{B}_H^{\prime 2}}
		\right) + \Or\left( \beta_H^2 \Mew^2 \right) .
	\end{equation}

	We must determine the vacuum polarization of the Higgs.
	As above, we assume this to follow from a conformal coupling to $\chi$.
	In the Standard Model, this would give the coupling
	\begin{equation}
		\fl
		S \supseteq - \frac{1}{2} \int \d^4 x \left\{
			B_H(\beta_H \chi) | (\partial_a + \im \vec{A}_a \cdot \vec{t}
			- \im B_a y) H |^2 - C_H(\beta_H \chi) \mu^2
			H^\dag H + \Or\Big( [H^\dag H]^2 \Big) \right\} ,
		\label{eq:Higgs-couple}
	\end{equation}
	where $\vec{A}$ and $B$ are the gauge fields of the unbroken
	$\sugroup(2)$ and $\ugroup(1)_Y$ symmetries, respectively;
	$\vec{t}$ are a set of appropriately normalized generators of
	$\sugroup(2)$; and $y$ is the generator of $\ugroup(1)$.
	$H$ is an $\sugroup(2)$ Higgs doublet,
	and $\mu$ is a standard parameter of the quartic Higgs potential,
	related to the Higgs mass by the rule $M_H^2 = 2 \mu^2$.
	In many models the phenomenological couplings $B$, $B_H$ and $C_H$
	will be closely related, but for the present we leave them arbitrary.
	If no relationship exists between the couplings, we find that
	unitarity is not respected
	at tree level in two-body scattering of gauge bosons
	\cite{LlewellynSmith:1973ey,Dicus:1992vj,Cornwall:1974km,Cornwall:1973tb}
	at energy scales above
	$ [G_F |B(\beta\bar{\chi}) - B_H(\beta_H\bar{\chi})|]^{-1/2}$.%
		\footnote{Additional violations of perturbative unitarity near
		the chameleon scale $\beta^{-1}$ may arise from new dark
		energy exchange diagrams in two-body 
		scattering.}
	In models containing more than one Higgs doublet
	we assume that Eq.~\eref{eq:Higgs-couple} continues to give
	a good approximation to the couplings of the lightest neutral Higgs.

	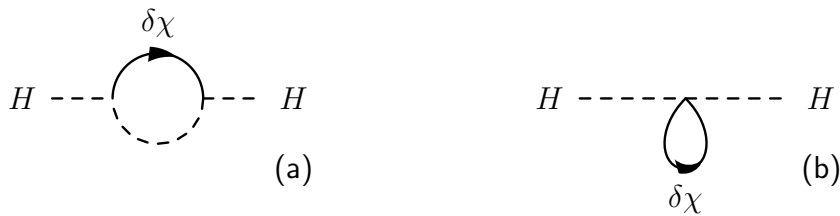
\begin{figure}
		\begin{center}
			\hfill
			\begin{fmfgraph*}(80,60)
				\fmfleft{l}
				\fmfright{r}
				\fmfpen{thin}
				\fmf{dashes,tension=3}{l,v1}
				\fmf{fermion,label=$\delta\chi$,left}{v1,v2}
				\fmf{dashes,left}{v2,v1}
				\fmf{dashes,tension=3}{v2,r}
				\fmfv{label=$H$}{l}
				\fmfv{label=$H$}{r}
			\end{fmfgraph*}
			\figlabel{a}
			\hspace{3mm}
			\hfill
			\hspace{3mm}
			\begin{fmfgraph*}(80,60)
				\fmfleft{l}
				\fmfright{r}
				\fmfpen{thin}
				\fmf{dashes,tension=2}{l,v}
				\fmf{fermion,label=$\delta\chi$,left}{v,v}
				\fmf{dashes,tension=2}{v,r}
				\fmfv{label=$H$}{l}
				\fmfv{label=$H$}{r}
			\end{fmfgraph*}
			\figlabel{b}
			\hfill
			\mbox{}
		\end{center}
		\caption{Processes contributing to the
			self-energy of the Higgs boson.
			An initial Higgs boson state, represented by a dashed line,
			radiates into scalar quanta $\chi$ (represented by a solid line)
			which are eventually re-absorbed
			to yield a final state characterized by the same quantum numbers
			and momentum as the initial state.
			\label{fig:polarizations}
		}
	\end{figure}

	\clearpage
	The Higgs vacuum polarization $\Pi_{HH}(k^2)$ can be computed.
	The one-loop contributions
	are shown in Fig.~\ref{fig:polarizations},
	and depend on the following vertices:
	\vspace{7mm}
	\begin{equation}
		\parbox{30mm}{
			\begin{fmfgraph*}(60,40)
				\fmfleft{l1,l2}
				\fmfright{r}
				\fmf{dashes}{l1,v}
				\fmf{dashes}{l2,v}
				\fmf{plain}{v,r}
				\fmfv{label=$k_2$}{l1}
				\fmfv{label=$k_1$}{l2}
				\fmfv{label=$k_3$}{r}
			\end{fmfgraph*}
			}
			\rightarrow
			\frac{\beta_H}{2} \Big\{
				\bar{B}'_H (k_2 \cdot k_3) -
				\bar{C}'_H M_H^2
			\Big\} ,
		\label{feynrule:w-cubic}
	\end{equation}
	\vspace{10mm}
	\begin{equation}
		\parbox{30mm}{
			\begin{fmfgraph*}(60,40)
				\fmfleft{l1,l2}
				\fmfright{r1,r2}
				\fmf{dashes}{l1,v}
				\fmf{dashes}{l2,v}
				\fmf{plain}{r1,v}
				\fmf{plain}{r2,v}
				\fmfv{label=$k_2$}{l1}
				\fmfv{label=$k_1$}{l2}
				\fmfv{label=$k_4$}{r1}
				\fmfv{label=$k_3$}{r2}
			\end{fmfgraph*}
			}
			\rightarrow
			\frac{\beta_H^2}{4} \Big\{
				\bar{B}''_H (k_2 \cdot k_3) -
				\bar{C}''_H M_H^2
			\Big\}  ,
		\label{feynrule:w-quartic}
	\end{equation}
	\vspace{7mm}

	The diagram in Fig.~\ref{fig:polarizations}\figlabel{a} corresponds to a
	vacuum polarization
	\begin{eqnarray}
		\fl\nonumber
		\Pi_{HH}(k^2)
		=
		\frac{\beta^2_H}{8\pi^2}
		\int_0^1 \; \d x
		\int_0^{\Lambda}
		\frac{\kappa^3 \; \d\kappa}{(\kappa^2 + \Sigma_H^2)^2}
		\\
		\mbox{} \times
		\left[
			\bar{B}^{\prime 2}_H \left(
				\frac{k^2\kappa^2}{4} + x^2 k^4
			\right)
			+
			\bar{C}_H^{\prime} M_H^2 (
				\bar{C}'_H M_H^2
				+ 2\bar{B}'_H k^2 x
			)
		\right] ,
	\end{eqnarray}
	where $x$ is a Feynman parameter and we have rotated to Euclidean
	signature.
	In analogy with Eq.~\eref{eq:SigmaZ},
	$\Sigma_H$ is defined so that
	\begin{equation}
		\Sigma_H^2 = x(1-x) k^2 + (1-x) M_H^2 + x M_\chi^2 .
	\end{equation}
	The diagram in Fig.~\ref{fig:polarizations}\figlabel{b} contributes
	\begin{equation}
		\Pi_{HH}(k^2)
		=
		-\frac{\beta_H^2}{16\pi^2}
		\int^{\Lambda}_0
		\frac{\kappa^3 \; \d\kappa}{\kappa^2 + M_{\chi}^2}
		(
			\bar{B}''_H k^2 + \bar{C}''_H M_H^2
		) .
	\end{equation}
	Carrying out the $\kappa$ integrals, we find
	\begin{eqnarray}
		\fl\nonumber
		\Pi_{HH}(k^2)
		=
		\frac{\beta_H^2}{8\pi^2}
		\frac{1}{\bar{B}_H}
			\int_0^1 \d x \;\Bigg\{
						\frac{\bar{B}_H^{\prime 2} k^2}{4}
				\left[
					\Lambda^2 -
					\frac{\Lambda^2}{2} \frac{\Lambda^2}{\Lambda^2 + \Sigma_H^2}
					- \Sigma_H^2 \ln \left(
						1 + \frac{\Lambda^2}{\Sigma_H^2}
					\right)
				\right]
				\\ \nonumber
				\hspace{-8mm}
				\mbox{} +
				\Big\{ B_H^{\prime 2} x^2 k^4
					+ M_H^2 \bar{C}_H'( 2 \bar{B}_H' x k^2
					+ M_H^2 \bar{C}_H') \Big\}
				\left[
					- \frac{1}{2} \frac{\Lambda^2}{\Lambda^2 + \Sigma_H^2}
					+ \frac{1}{2} \ln \left(
						1 + \frac{\Lambda^2}{\Sigma_H^2}
					\right)
				\right]
				\\
				\hspace{-8mm}
				\mbox{} -
				\frac{\bar{B}_H}{2}
				(
					\bar{B}''_H k^2 + M_H^2 \bar{C}''_H
				)
				\left[
					\frac{\Lambda^2}{2} - \frac{M_\chi^2}{2}
					\ln \left(1 + \frac{\Lambda^2}{M_\chi^2} \right)
				\right]
			\Bigg\} .
		\label{eq:Higgs-vac}
	\end{eqnarray}
	From Eqs.~\eref{eq:self-energy} and~\eref{eq:Higgs-vac} it is possible
	to compute $R$, the parameter which summarizes the oblique dependence
	of the rates $\Gamma(ZZ \rightarrow h)$ and
	$\Gamma( W^+ W^- \rightarrow h)$.
	We find
	\begin{equation}
		\fl
		\alpha R = \frac{\beta_H^2 \Lambda^2}{32\pi^2}
		\frac{\bar{B}_H^{\prime 2}}{\bar{B}}
		\left[
			\frac{1}{2} \left( 1 + \frac{\bar{B}}{\bar{B}_H} \right)
			-
			2 \frac{\bar{B}_H'' \bar{B}}{\bar{B}_H^{\prime 2}}
		\right] +
		\mbox{finite terms of order $\Or\left( \beta_H^2 \Mew^2 \right)$} .
		\label{eq:R}
	\end{equation}
	At leading order in the divergence, it is independent of $C_H$.
	We note, however, that a dependence on $C_H$ persists
	among those terms which are finite in
	the limit $\Lambda \rightarrow \infty$.
	These finite terms are of order $\beta_H^2 \Mew^2$ and
	can be neglected when $\beta_H \ll (1\;\mbox{TeV})^{-1}$.
	We conclude that the oblique correction is very small, unless the
	divergent part of Eq.~\eref{eq:R} can contribute a significant effect.

	What is the meaning of the divergent term in Eq.~\eref{eq:R}?
	One must be wary when reasoning with power-law divergences, because
	they can be ascribed no invariant significance. For example,
	they are absent in dimensional regularization.
	In Ref.~\cite{Brax:2009aw}
	it was found that similar divergences could
	be absorbed in renormalizations
	of $G_F$ and the vector boson masses, $M_Z$ and $M_W$.
	The divergence in Eq.~\eref{eq:R} cannot be absorbed in this way.
	It expresses a sensitivity to whatever physics completes
	the low-energy theory containing the Standard Model
	and the effective interaction,
	Eq.~\eref{eq:chi-gauge-effective-operator}.
	It is not a prediction that large effects 
	should be observed in a particle collider.
	The same divergence arises in all models with this low-energy limit,
	irrespective of what physics takes place at high energy.
	To fix its value with confidence, we must know the details of
	the high energy completion.
	In the next section we compute its value for a model in which
	the low energy theory is obtained by integrating out heavy fermions
	with SUSY-like couplings.
	
	\subsection{Higgs couplings from heavy charged particles}
	\label{sec:charginos-higgs}
	If it is obtained as the low energy effective theory of some
	complete UV physics,
	Eq.~\eref{eq:chi-gauge-effective-operator}
	will be accompanied by other interactions
	which cannot be neglected.
	The most important is a new contact interaction between
	gauge bosons and the Higgs field, and in what follows we
	estimate its effect on Higgs production.
	We use the size of the contribution to Higgs production from
	this straight corrections to estimate the size of the cutoff
	that controls corrections due to interactions with dark energy.
	
	Gauge invariance
	constrains
	which operators can appear in the low-energy theory.
	In the minimal scenario we are considering,
	the Higgs field is in the fundamental
	$\bf{2}$ representation of $\sugroup(2)$.
	By construction, a field strength term such as
	$F^{ab} F_{ab}$ transforms in the adjoint representation.
	Therefore the lowest order non-trivial interaction
	with the Higgs
	must involve
	$H^\dag H$,
	making $H^\dag H \Tr( F^{ab} F_{ab} )$ a gauge invariant
	dimension-six operator.	
	
	If this operator is present in the low-energy effective theory,
	it will modify our expectation for Higgs production.
	Accordingly, we must evaluate its coefficient.
	The prediction is model-dependent.
	Consider a minimal scenario, where the heavy charged particles
	are fermions which have vertices with the lightest neutral Higgs
	of the form
	\vspace{7mm}
	\begin{equation}
		\parbox{30mm}{
			\begin{fmfgraph*}(60,40)
				\fmfleft{l}
				\fmfright{r1,r2}
				\fmf{dashes}{l,v}
				\fmf{fermion}{r1,v}
				\fmf{fermion}{v,r2}
				\fmfv{label=$h$}{l}
				\fmfv{label=$\chi_j$}{r1}
				\fmfv{label=$\chi_i$}{r2}
			\end{fmfgraph*}
			}
			\rightarrow
			\im \coupling
			\left( C_{ij} L + C_{ji}^\ast R \right) ,
		\label{eq:chargino-couplings}
	\end{equation}
	\vspace{7mm}
	where $i$ and $j$ label different species of fermion and
	the strength of the coupling is parametrized by $g$.
	The operator
	$L \equiv (1 + \gamma_5)/2$ projects onto the left-chirality half
	of a Dirac spinor, and $R \equiv (1 - \gamma_5)/2$ is its conjugate.
	The $C_{ij}$ should be chosen real and symmetric if CP violation
	is to be avoided.
	
	In a supersymmetric Standard Model, the $\chi_i$ will be charginos
	and neutralinos. These have couplings to the lightest neutral
	Higgs of the form~\eref{eq:chargino-couplings},
	with $C_{ij}$ determined by the various factors which
	diagonalize the chargino and neutralino mass matrices.
	Explicit expressions can be found in {\S}A.7 of Ref.~\cite{Gunion:1989we}.
	In this case, the chargino and neutralino loops would be accompanied
	by heavy slepton loops which we do not calculate.
	There is no reason to expect the slepton contribution to be larger
	than the neutralino or chargino terms, so we anticipate that the
	fermion contribution alone is representative.
		
\begin{figure}
		\vspace{5mm}
		\mbox{}
		\hspace{15mm}
		\begin{fmfgraph*}(110,50)
			\fmfleft{l1,l2}
			\fmfright{r1,r2}
			\fmf{boson}{l1,v1}
			\fmf{boson}{l2,v2}
			\fmf{dashes}{r1,v3}
			\fmf{dashes}{r2,v4}
			\fmf{dbl_plain,label=$k$,label.side=left}{v2,v1}
			\fmf{dbl_plain,label=$k+q$,label.side=left}{v1,v3}
			\fmf{dbl_wiggly,label=$k+q+r$,label.side=left}{v3,v4}
			\fmf{dbl_plain,label=$k+q+r+s$,label.side=left}{v4,v2}
			\fmfv{label=$A_a(q)$,label.angle=180}{l1}
			\fmfv{label=$A_b(p)$,label.angle=180}{l2}
			\fmfv{label=$H(r)$,label.angle=0}{r1}
			\fmfv{label=$H(s)$,label.angle=0}{r2}
			\fmffreeze
			\fmfforce{(0,0)}{l2}
			\fmfforce{(0,h)}{l1}
			\fmfforce{(w,0)}{r2}
			\fmfforce{(w,h)}{r1}
			\fmfforce{(0.35w,0)}{v2}
			\fmfforce{(0.35w,h)}{v1}
			\fmfforce{(0.65w,0)}{v4}
			\fmfforce{(0.65w,h)}{v3}
		\end{fmfgraph*}
		\hfill
		\begin{fmfgraph*}(110,50)
			\fmfleft{l1,l2}
			\fmfright{r1,r2}
			\fmf{boson}{l1,v2}
			\fmf{boson}{l2,v1}
			\fmf{dashes}{r1,v3}
			\fmf{dashes}{r2,v4}
			\fmf{dbl_plain,label=$k$,label.side=right}{v2,v1}
			\fmf{dbl_plain,label=$k+q$,label.side=left}{v1,v3}
			\fmf{dbl_wiggly,label=$k+q+r$,label.side=left}{v3,v4}
			\fmf{dbl_plain,label=$k+q+r+s$,label.side=left}{v4,v2}
			\fmfv{label=$A_a(q)$,label.angle=180}{l1}
			\fmfv{label=$A_b(p)$,label.angle=180}{l2}
			\fmfv{label=$H(r)$,label.angle=0}{r1}
			\fmfv{label=$H(s)$,label.angle=0}{r2}
			\fmffreeze
			\fmfforce{(0,0)}{l2}
			\fmfforce{(0,h)}{l1}
			\fmfforce{(w,0)}{r2}
			\fmfforce{(w,h)}{r1}
			\fmfforce{(0.35w,0)}{v2}
			\fmfforce{(0.35w,h)}{v1}
			\fmfforce{(0.65w,0)}{v4}
			\fmfforce{(0.65w,h)}{v3}
		\end{fmfgraph*}
		\hspace{15mm}
		\mbox{}
		
		\vspace{20mm}
		\mbox{}
		\hspace{15mm}
		\begin{fmfgraph*}(110,50)
			\fmfleft{l1,l2}
			\fmfright{r1,r2}
			\fmf{boson}{l1,v1}
			\fmf{boson}{l2,v2}
			\fmf{dashes}{r1,v4}
			\fmf{dashes}{r2,v3}
			\fmf{dbl_plain,label=$k$,label.side=left}{v2,v1}
			\fmf{dbl_plain,label=$k+q$,label.side=left}{v1,v3}
			\fmf{dbl_wiggly}{v3,v4}
			\fmf{dbl_plain,label=$k+q+r+s$,label.side=left}{v4,v2}
			\fmfv{label=$A_a(q)$,label.angle=180}{l1}
			\fmfv{label=$A_b(p)$,label.angle=180}{l2}
			\fmfv{label=$H(r)$,label.angle=0}{r1}
			\fmfv{label=$H(s)$,label.angle=0}{r2}
			\fmffreeze
			\fmfforce{(0,0)}{l2}
			\fmfforce{(0,h)}{l1}
			\fmfforce{(w,0)}{r2}
			\fmfforce{(w,h)}{r1}
			\fmfforce{(0.35w,0)}{v2}
			\fmfforce{(0.35w,h)}{v1}
			\fmfforce{(0.65w,0)}{v4}
			\fmfforce{(0.65w,h)}{v3}
		\end{fmfgraph*}
		\hfill
		\begin{fmfgraph*}(110,50)
			\fmfleft{l1,l2}
			\fmfright{r1,r2}
			\fmf{boson}{l1,v2}
			\fmf{boson}{l2,v1}
			\fmf{dashes}{r1,v4}
			\fmf{dashes}{r2,v3}
			\fmf{dbl_plain,label=$k$,label.side=right}{v2,v1}
			\fmf{dbl_plain,label=$k+q$,label.side=left}{v1,v3}
			\fmf{dbl_wiggly}{v3,v4}
			\fmf{dbl_plain,label=$k+q+r+s$,label.side=left}{v4,v2}
			\fmfv{label=$A_a(q)$,label.angle=180}{l1}
			\fmfv{label=$A_b(p)$,label.angle=180}{l2}
			\fmfv{label=$H(r)$,label.angle=0}{r1}
			\fmfv{label=$H(s)$,label.angle=0}{r2}
			\fmffreeze
			\fmfforce{(0,0)}{l2}
			\fmfforce{(0,h)}{l1}
			\fmfforce{(w,0)}{r2}
			\fmfforce{(w,h)}{r1}
			\fmfforce{(0.35w,0)}{v2}
			\fmfforce{(0.35w,h)}{v1}
			\fmfforce{(0.65w,0)}{v4}
			\fmfforce{(0.65w,h)}{v3}
		\end{fmfgraph*}
		\hspace{15mm}
		\mbox{}
		\vspace{5mm}
		\caption{Leading interactions between the Higgs field
		and electroweak gauge bosons. The interactions are mediated by
		two species of chargino, $\chargino$ (straight doubled lines)
		and $\chargino'$ (wiggly doubled lines), of masses
		$M$ and $M'$, respectively.
		\label{fig:higgs-diagrams}}
	\end{figure}
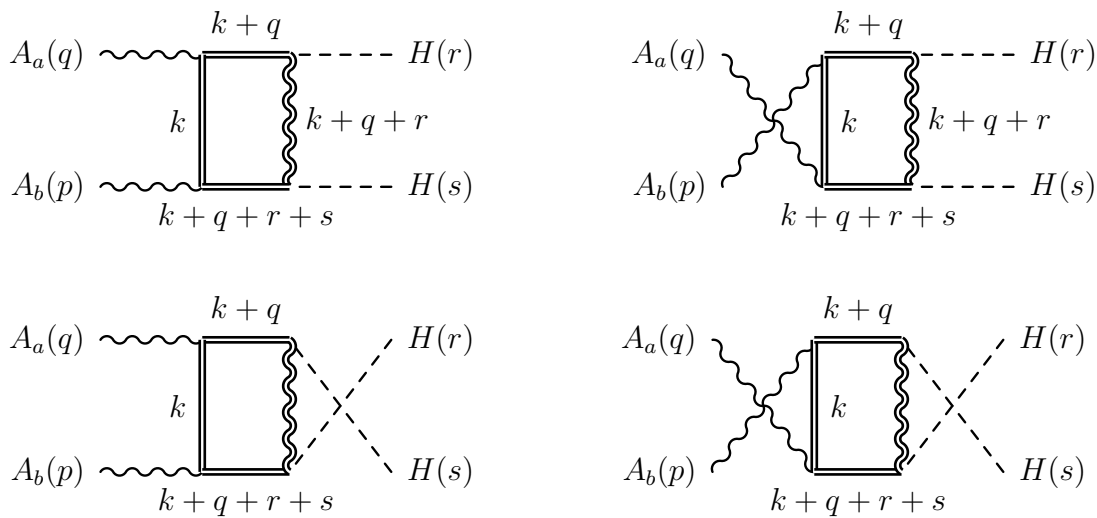

	Eq.~\eref{eq:chargino-couplings} gives rise to effective operators
	depicted in Fig.~\ref{fig:higgs-diagrams}.
	We denote a fermion of species $i$ by a straight doubled line,
	and species $j$ by a wiggly doubled line.
	These diagrams must be summed over all $i$ and $j$.
	The correlation function
	$\langle A_a(q) A_b(p) h(r) h(s) \rangle$ satisfies
	\begin{eqnarray}
		\fl\nonumber
		\langle A_a(q) A_b(p) h(r) h(s) \rangle =
		(2\pi)^4 \delta(p + q + r + s)
		\egauge^2 \coupling^2 h(r) h(s) A_a(q) A_b(p)
		\int \frac{\d^4 k}{(2\pi)^4} \\ \nonumber
		\mbox{} \times
		\tr \sum_{ij} \Bigg\{
			\gamma^a
			\frac{-\im(\slashed{k} + \slashed{q}) + M_i}
				{(k+q)^2 + M_i^2 - \im \epsilon}
			(C_{ij} L + C_{ji}^\ast R)
			\frac{-\im(\slashed{k} + \slashed{q} + \slashed{r}) + M_j}
				{(k+q+r)^2 + M_j^2 - \im \epsilon}
			\\ \nonumber
			\hspace{12mm}
			\mbox{} \times
			(C_{ji} L + C_{ij}^\ast R)
			\frac{-\im(\slashed{k} + \slashed{q} + \slashed{r} + \slashed{s})
				+ M_i}
				{(k + q + r + s)^2 + M_i^2 - \im \epsilon}
			\gamma^b
			\frac{-\im \slashed{k} + M_i}
				{k^2 + M_i^2 - \im \epsilon}
		\Bigg\} \\
		\mbox{}
		+ \Big(
			\begin{array}{c} p \leftrightarrow q \\ a \leftrightarrow b
			\end{array}
		\Big)
		+ ( r \leftrightarrow s ) ,
		\label{eq:higgs-gauge-correlator}
	\end{eqnarray}
	where $M_i$ is the mass of species $i$
	and the symmetrizations are nested, so that
	$r \leftrightarrow s$ is carried out after the simultaneous
	exchanges $p \leftrightarrow q$ and $a \leftrightarrow b$.
	As above, the $k$ integral is divergent and potentially dependent
	on the labeling of momenta inside the loop, but
	once the integral has been regularized
	it can be checked that this dependence vanishes.
	
	After a tedious calculation, we find that
	Eq.~\eref{eq:higgs-gauge-correlator}
	can be reproduced using the following effective Lagrangian,
	\begin{eqnarray}
		\fl\nonumber
		\Leff \supseteq
		\sum_{ij}
		\frac{\coupling^2}{3!(4\pi)^2}
		\frac{2}{18 M_i^2 (1 - x_{ij}^2)^5}
		\Big\{
			& \egauge^2	\zeta_1(x_{ij}) H^\dag H \Tr F^{ab} F_{ab}
			+ \zeta_2(x_{ij}) |D^2 H|^2
			\\ &
			\mbox{}
			+ \zeta_3(x_{ij}) (D_a D_b H)^\dag (D^a D^b H)
		\Big\}
		\label{eq:higgs-gauge-effective-operator}
	\end{eqnarray}
	where $x_{ij} \equiv M_j/M_i$
	and the functions $\{ \zeta_1, \zeta_2, \zeta_3 \}$
	satisfy
	\begin{eqnarray}
		\fl\nonumber
		\zeta_1(x_{ij}) \equiv
		\Sigma_{ij} \left[
			8 - 49 x_{ij}^2 + 99 x_{ij}^4 - 71 x_{ij}^6
			+ 13 x_{ij}^8 - 12 x_{ij}^4 (
				3 - 7 x_{ij}^2 + 2 x_{ij}^4
			) \ln x_{ij}
		\right] \\
		\hspace{-9mm} \mbox{}
		- 2 x_{ij} \Gamma_{ij} \left[
			4 - 6  x_{ij}^2 - 9 x_{ij}^4 + 14 x_{ij}^6 - 3 x_{ij}^8
			+ 6 x_{ij}^2 (
				3 - 6 x_{ij}^2 + x_{ij}^4
			) \ln x_{ij}
		\right] , \\
		\fl\nonumber
		\zeta_2(x_{ij}) \equiv
		2 \Sigma_{ij} \left[
			7 - 47 x_{ij}^2 + 63 x_{ij}^4 - 25 x_{ij}^6 + 2 x_{ij}^8
			- 12 x_{ij}^4 (
				6 - 5 x_{ij}^2 + x_{ij}^4
			) \ln x_{ij}
		\right] \\
		\hspace{-9mm} \mbox{}
		- 4x_{ij} \Gamma_{ij} \left[
			2 + 9 x_{ij}^2 - 18 x_{ij}^4 + 7 x_{ij}^6 + 6 x_{ij}^2 (
				3 - x_{ij}^4
			) \ln x_{ij}
		\right] , \\
		\fl\nonumber
		\zeta_3(x_{ij}) \equiv
		-2 \Sigma_{ij} \left[
			4 - 23 x_{ij}^2 + 63 x_{ij}^4 - 49 x_{ij}^6 + 5 x_{ij}^8
			+12 x_{ij}^6 (
				5 - x_{ij}^2
			) \ln x_{ij}
		\right] \\
		\hspace{-9mm} \mbox{}
		+ 2x_{ij} \Gamma_{ij} \left[
			1 - 9 x_{ij}^2 - 9 x_{ij}^4 + 17 x_{ij}^6 - 12 x_{ij}^4 (
				3 + x_{ij}^2
			) \ln x_{ij}
		\right] .
	\end{eqnarray}
	The quantities $\Sigma_{ij}$ and $\Gamma_{ij}$ are defined by
	\begin{eqnarray}
		\Sigma_{ij} \equiv C_{ij} C_{ij}^\ast + C_{ji} C_{ji}^\ast
		\hspace{50mm} \mbox{\textsf{(no sum on $i$ or $j$)}} \\
		\Gamma_{ij} \equiv C_{ij} C_{ji} + C_{ij}^\ast C_{ji}^\ast
		\hspace{50.5mm} \mbox{\textsf{(no sum on $i$ or $j$)}}
	\end{eqnarray}
	To exhibit its gauge invariance,
	Eq.~\eref{eq:higgs-gauge-effective-operator} has been written in terms
	of a conventionally normalized $\sugroup(2)$ doublet $H$,
	which coincides with the field-space orientation of the lightest
	neutral Higgs.
	It will be accompanied by higher derivative terms, represented by
	$\Or(\partial^4)$, which have been neglected.
	There is also a term of the form
	$c h^2 A_a A^a$, with $c$ a divergent constant, whose role is
	to renormalize the charge $\egauge$.
	We discard this term and take $\egauge$ to be the renormalized charge.
	At higher order in $H$, arbitrary powers of $H^\dag H$ may be
	generated. These will lead to higher-dimension
	operators which couple polynomials of $H$ and its derivatives to the
	gauge field, but at leading order we can restrict our attention to
	Eq.~\eref{eq:higgs-gauge-effective-operator}.

	To determine the correct order of magnitude for
	$\Lambda$, the cutoff used in the calculation of the oblique
	corrections in \S\ref{sec:low-energy-oblique},
	it is safest to
	match to whatever theory controls physics in the ultra-violet
	\cite{Burgess:1992gx}.
	In the present case, this is summarized by
	Eq.~\eref{eq:higgs-gauge-effective-operator}.	
	Using the effective interactions in
	Eq.~\eref{eq:higgs-gauge-effective-operator}
	it is possible to compute the enhancement to Higgs production
	due to these straight corrections and the represent this in
	the form of an oblique correction.	This will give an estimate
	of the size of the cut off needed in the calculation of the
	oblique corrections in	\S\ref{sec:low-energy-oblique}.
	As a reasonable approximation, we take the $\sugroup(2)$
	doublet, $H$, to develop a vacuum expectation value of order
	the Standard Model scale, $G_F^{-1/2}$.
	We take neutral excitations around this condensate
	to be representative of the interaction with the lightest neutral Higgs.
	Choosing the gauge field to be the
	Abelian vector associated with the $Z$,
	the resulting effective Lagrangian is
	\begin{eqnarray}
		\fl\nonumber
		\Leff \supseteq \frac{\egauge^2 \coupling^2}{2592 \pi^2}
		\sum_{ij} \frac{(\sqrt{2} G_F)^{-1}}{M_i^2 (1-x_{ij}^2)^5}
		\Big\{
			[2 \zeta_1(x_{ij}) + 2 \zeta_3(x_{ij})] h
			(\partial_a Z_a \partial^a Z^b -
			\partial_a Z_b \partial^b Z^a)
			\\ \hspace{3.4cm} \mbox{}
			-
			[ \zeta_2(x_{ij}) + \zeta_3(x_{ij}) ]
			\partial^2 h Z^a Z_a
			+
			\zeta_3(x_{ij}) h Z^b \partial^2 Z_b
		\Big\} .
	\end{eqnarray}
	In the limit where the relative velocity of the colliding
	vector bosons goes to zero,%
		\footnote{The calculation does not need to be restricted to this
		kinematic limit, but it leads to simpler final expressions.
		Since the vector bosons are taken to be on-shell in the
		effective $W$ approximation, the invariant magnitude of
		any momenta will be of order $\sim M_W$.
		This implies that although the result may be modified by
		factors of $\Or(1)$, it is unlikely that we commit a gross error by
		specializing to the zero-velocity limit.}
	and defining
	\begin{equation}
		\hat{\zeta}_m \equiv \sum_{ij} \frac{\zeta_m(x_{ij})}
			{M_i^2 (1 - x_{ij}^2)^5} ,
	\end{equation}
	we find
	\begin{equation}
		\fl
		\frac{\delta \Gamma(ZZ \rightarrow h)}
			{\Gamma(ZZ \rightarrow h)}
		=
		\left(
			\frac{\egauge^2 \coupling^2}{2592 \sqrt{3} \pi^2 G_F}
		\right)^2
		\Big\{
			28 \hat{\zeta}_1^2 - 32 \hat{\zeta}_1 \hat{\zeta}_2
			- 28 \hat{\zeta}_1 \hat{\zeta}_3 + 25 \hat{\zeta}_2 \hat{\zeta}_3
			+ 35 \hat{\zeta}_3^2
		\Big\} .
		\label{eq:higgs-rate-change}
	\end{equation}
	Despite appearances Eq.~\eref{eq:higgs-rate-change} is dimensionless,
	because each $\hat{\zeta}_i$ has the same dimension
	as $G_F$, [mass]$^{-2}$.
	The magnitude of this correction varies with the mass ratio,
	$x_{ij}$, approaching zero as $x_{ij} \rightarrow 1$
	but asymptoting to an approximate constant for large or small ratios.
	In a typical supersymmetric Standard Model the chargino and
	neutralino masses are undetermined, but provided there is not total
	mass degeneracy we expect that this threshold correction generates
	a contribution represented by a cutoff of order
	\begin{equation}
		\Lambda
		\simeq
		\beta_H^{-1}
		\times
		\frac{\egauge^2 \coupling^2}{M^2 G_F} .
		\label{eq:cutoff}
	\end{equation}
	This is typically a rather small number, leading to an essentially
	negligible enhancement in the Higgs production rate.

	\section{Conclusions}
	\label{sec:conclusions}

	In this paper we have derived
	the low energy effective theory which governs interactions between
	the gauge bosons of the electroweak sector and a dark energy scalar field.
	The dark energy is taken to have conformal couplings to the matter
	species of the Standard Model.
	It is possible that couplings of this type allow so-called ``chameleon''
	behaviour, in which the field dynamically adjusts its mass
	to be large in regions of high average density, and small elsewhere.
	If such theories exist then they would lead to an unambiguous
	prediction of light dark energy quanta interacting in the beam pipe
	of any particle accelerator.
	Our low energy theory applies
	strictly in any model containing heavy charged
	fermions, but a very similar effective Lagrangian would apply for
	a model containing heavy charged particles of any spin.
	As a specific example, any supersymmetric Standard Model must contain
	charginos and neutralinos. These carry the quantum numbers of the
	$\sugroup(2) \times \ugroup(1)$ gauge group and have masses of order
	the supersymmetry-breaking scale.
	However, our calculation is not restricted to the supersymmetric case.
	
	One might worry that the presence of chameleonic quanta would change
	our predictions for the outcome of particle physics experiments.
	In Ref.~\cite{Brax:2009aw} we argued this did not happen for
	any process without Higgs quanta.
	In this paper we have extended our argument to include Higgs
	production.
	In particular,
	using the low energy theory
	we have computed the oblique corrections
	to each of the Higgs, $Z$ and $W^\pm$ propagators
	at energies below the mass, $M$, of the heavy charged fermions.
	Such corrections modify the rates
	$\Gamma(ZZ \rightarrow h)$
	and
	$\Gamma(W^+ W^- \rightarrow h)$,
	where $h$ is the lightest neutral Higgs,
	and in principle could change the rate of production of this particle at
	a hadron collider. We find that the corrections diverge
	quadratically with the scale chosen as the cut off for the
	effective theory. This does not predict a large enhancement to
	the production of Higgs bosons from interactions with dark
	energy, instead indicating that the interactions with dark
	energy make this process sensitive to the UV physics.
	
	Other contributions exist, generated by processes taking place at
	high energy, which are integrated out of the low-energy
	description.
	We determine these ``threshold corrections'' by
	integrating out heavy fermion loops which mediate interactions
	between the lightest neutral Higgs and the gauge bosons.  The
	scale of these corrections can then be reinterpreted as	 a cutoff
	in an oblique calculation, of order
	$\Lambda \sim \beta_H^{-1} \egauge^2 \coupling^2 / M^2 G_F^2$,
	and therefore leads to at most small effects.
	When $M$ is much larger than the Standard Model scale $G_F^{-1/2}$
	it is entirely negligible.
	In an unconstrained theory, we might have anticipated a cutoff
	of order $\Lambda \sim \beta_H^{-1}$, because
	at this scale
	the effective dark energy theory becomes invalid.
	If this were true, it would be possible to contemplate
	corrections to the Higgs production rate of $\Or(1)$ or larger,
	which would lie within the discovery reach of the LHC.

	Unfortunately, the corrections
	are much smaller.
	Large terms could only arise from the relevant operator
	$H Z^a Z_a$, but its appearance is forbidden by gauge invariance
	above the scale of electroweak symmetry breaking.
	Therefore, we expect the coefficient of this term
	to be at most $G_F^{-1/2}$, rather than $\beta_H^{-1}$.
	Instead, the leading correction comes from the operators
	$H^\dag H \Tr F^{ab} F_{ab}$,
	$|D^2 H|^2$
	and $(D_a D_b H)^\dag (D^a D^b H)$.
	These are dimension-six operators,
	because of the $\sugroup(2)$ nature of the Higgs doublet.
	Accordingly the cutoff is suppressed by $(M^2 G_F)^{-1}$, making
	it small. This must be typical of any UV correction because
	there is no relevant operator we can write down which will
	couple $\chi$ to the gauge fields.	To get a larger effect, it
	appears to be necessary to break the gauge invariance of the
	theory. This does not rule out the possibility that dark energy could
	be responsible for an enhancement in the Higgs production rate, but such
	a scenario would apparently require exotic physics.

 	In the context of particle colliders where strong magnetic fields
	are present, the coupling derived in \S\ref{sec:axionic} implies a
	coupling of the
	chameleon to synchrotron radiation.
	This would lead to emergence of
	chameleon-like particles due to the Primakov effect. For most
	chameleon theories, the large mass assumed by chameleon-like
	particles in a dense environment implies that the beam pipe acts
	as a reflecting wall, preventing dark energy particles from
	leaking out. Hence collider experiments would actually take place
	in a dark energy bath. The analysis of this phenomenon is left for
	future work.
	
	\ack
	We would like to thank Ben Allanach, Neil Barnaby,
	Cliff Burgess, Douglas Shaw and James Stirling for helpful discussions.
	CB is supported by the German Science Foundation (DFG) under the
	Collaborative Research Centre (SFB) 676.
	ACD and DS are supported by STFC and the Cambridge Centre for
	Theoretical Cosmology.
	AW is supported by the Cambridge Centre for Theoretical Cosmology.

	\section*{References}

\providecommand{\href}[2]{#2}\begingroup\raggedright\endgroup

\end{fmffile}
\end{document}